\documentclass[
final,%
twocolumn,
]%
{revtex4}

\usepackage{graphicx}
\usepackage{dcolumn}
\usepackage{bm}
\usepackage{subfigure}
\usepackage{amssymb}
\usepackage{amsmath}

\def\be{\begin{equation}}
\def\ee{\end{equation}}
\def\e#1{\label{#1}\end{equation}}
\def\r#1{(\ref{#1})}
\def\bem#1{\begin{mathletters}\label{#1}}
\def\eml{\end{mathletters}}

\def\ket#1{{|#1\rangle}}
\def\bra#1{{\langle#1|}}

\def\mean#1{{\langle#1\rangle}}
\def\4#1{{\boldsymbol{#1}}}
\def\8#1{{\widetilde{#1}}}
\def\bse{\begin{subequations}}
\def\ese{\end{subequations}}
\def\eqref#1{(\ref{#1})}
\def\sinc{{\rm sinc}}

\begin{document}

\title{Non-Markovian control of qubit thermodynamics by frequent quantum measurements}

\author{Guy Bensky,
D. D. Bhaktavatsala Rao,
Goren Gordon,
David Gelbwaser-Klimovsky,
Noam Erez,
Gershon Kurizki}
\affiliation{%
Department of Chemical Physics, Weizmann Institute of
Science,P. O. Box 26, Rehovot 76100, Israel\\
Tel: +972-8-934-3918, Fax: +972-8-934-4123}%

\begin{abstract}
We explore the effects of frequent, impulsive quantum nondemolition
measurements of the energy of two-level systems (TLS), alias qubits,
in contact with a thermal bath. The resulting entropy and temperature
of both the system and the bath are found to be completely determined
by the measurement rate, and unrelated to what is expected by standard
thermodynamical rules that hold for Markovian baths. These anomalies
allow for {\em very fast control} of heating, cooling and
state-purification (entropy reduction) of qubits, much sooner than
their thermal equilibration time.
\end{abstract}

\maketitle
\section{Introduction}
{\em Non-Markovian quantum thermodynamics} of two level systems (TLS)
in contact with a bath has surprising aspects in store. According to
standard Markov thermodynamics, the TLS (alias qubit) thermal
equilibration process is expected to progress monotonically,
accompanied by increase of the entropy, at least on average
\cite{landau1980spe,spo78,ali79,jar97,lin74}. Yet drastic deviations
from this trend are revealed when considering impulsive disturbances
of thermal equilibrium between TLS and a bath
\cite{schulman2006rue,piilo2007qbm}. These effects bear certain
similarities to the work described in \cite{Nieu02}. We have shown
\cite{ere08Nature} that frequent and brief quantum non demolition
(QND) measurements of the TLS energy-states entail unfamiliar
anomalies of the entropy and temperature of both system and bath,
which become unrelated to what is known from standard, Markovian
thermodynamic rules\cite{spo78,lin74}: (i) a transition {\em from
heating to cooling} of the TLS ensemble as we vary the interval
between consecutive measurements on the time scale of the inverse
energy separation of the qubit levels; and (ii) correspondingly, {\em
oscillations} of the entropy relative to that of the equilibrium
state.

Here we present an in-depth study of short-time evolution of quantum
systems coupled to a bath, interrupted by frequent measurements.  We
first discuss in Sec.~\ref{sec-general} the initial equilibrium state
relevant to our scenario. Sec.~\ref{sec-disturbance} then describes
the measurement-induced disturbance of equilibrium. In
Sec.~\ref{sec-main-text} we present a master equation analysis of the
post-measurement evolution and a discussion of the heating and cooling
requirements. Cooling conditions and entropy evolution of the system
are discussed in Sec.~\ref{sec-cooling-cond} and ~\ref{sec-supp-D},
respectively. A discussion of possible experimental realizations is
given in Sec.~\ref{sec-conc}.

\section{System-bath entanglement at equilibrium}
\label{sec-general}
\subsection{Hamiltonian}
The following Hamiltonian describes the qubit system that interacts
with the bath.
\be
\label{H_total}
H_{tot}=H_S+H_B+H_{SB}.
\ee
Here $H_{tot}$ pertains to the coupled system and bath and consists of:
\begin{align}
\label{H-S}
H_S =&\hbar\omega_a\ket{e}\bra{e}, \\
\label{H-B}
H_B =&\hbar\sum_\lambda\omega_\lambda a_\lambda^\dagger a_\lambda,  \\
\label{H-I}
H_{SB} =&\mathcal{SB}, \mathcal{S} = \sigma_x,~~
\mathcal{B} = \hbar \sum_\lambda\left(\kappa_\lambda a_\lambda+\kappa_\lambda^* a_\lambda^\dagger\right),
\end{align}
where $\mathcal{S}$ and $\mathcal{B}$ are the system and bath
operators, respectively, in the system-bath interaction $H_{SB}$,
$a_\lambda(a^\dagger_\lambda)$ are the annihilation (creation)
operators, and $\kappa_\lambda$ is the matrix element of the weak
coupling to bath mode $\lambda$. We stress that in the interaction
Hamiltonian ($H_{SB}$) we {\em do not} invoke the rotating-wave
approximation (RWA)\cite{coh92}, namely, we do not impose energy
conservation between the system and the bath, on the time scales
considered\cite{kof04}.

\subsection{Qubit state mixedness at equilibrium}
At equilibrium, the qubit and the bath are in an entangled state. To
find the mean energy mixedness (impurity) of the qubit (TLS) at a
given temperature $T$, one needs the equilibrium density matrix for
the total system $\rho_{Eq} = \exp(-\beta H_{tot})/Z$, where $Z$ is
the partition function and $1/\beta = k_BT$.

Using Heims perturbation theory \cite{Heims} one can expand $\rho_{Eq}$ as
\begin{multline}
\rho_{Eq} =\frac{1}{Z}{\rm e}^{-\beta(H_0 + H_{SB})}\\
 =\frac{1}{Z}{\rm e}^{-\beta H_0}[1+\epsilon S_1 + \epsilon^2 S_2 + O(\epsilon^3)+\cdots],
\end{multline}
where $$\epsilon = \max(\eta_k/\hbar\omega_a),$$ is a small
dimensionless parameter normalizing the rate $\eta_k$ of the maximally
coupled mode to the TLS natural frequency, and
\begin{subequations}
\be
\epsilon S_1 = \beta\int^{1}_0 dx {\rm e}^{x\beta  H_0}H_{SB}{\rm e}^{-x\beta H_0},
\ee
\begin{multline}
\epsilon^2 S_2 =\\
\beta^2\int^1_0dx\int^x_0 dy {\rm e}^{x\beta H_0}H_{SB}{\rm e}^{-(x-y)\beta H_0}H_{SB}{\rm e}^{-y\beta H_0}.
\end{multline}
\end{subequations}
Noting that ${\rm e}^{-\beta H_0} = Z_0\rho_S\otimes\rho_B$, where $\rho_S$ and $\rho_B$ are the equilibrium density matrices for the system and the bath without interaction, the trace over the bath degrees of freedom can be performed. 
The state of the system is diagonal in the $\sigma_z$ basis and is given by 
$$\rho_S(\epsilon) = \frac{1}{2}(\mathcal{I}+P_{Eq}(\epsilon) \sigma_z).$$
The qubit purity at equilibrium is given by
\begin{multline}
\label{hp}
P_{Eq}(\epsilon) = \\
\frac{P_{Eq}+\epsilon^2\int^\infty_{-\infty}d\omega G_T(\omega)\left[P^+_{Eq}\mathcal{K}^+(\omega) - P^-_{Eq}\mathcal{K}^-(\omega) \right]}{1+\epsilon^2\int^\infty_{-\infty}d\omega G_T(\omega)\left[P^+_{Eq}\mathcal{K}^+(\omega) + P^-_{Eq}\mathcal{K}^-(\omega) \right] }.
\end{multline}
Here the temperature-dependent coupling spectrum 
\begin{subequations}
\be
G_T(\omega) = G_0(\omega)(n_T(\omega)+1)+G_0(-\omega)n_T(-\omega),
\ee
is written in terms of the
average occupation number at inverse temperature $\beta=1/T$,
\be
n_T(\omega) = \frac{1}{\exp(\beta \hbar\omega)-1},
\ee
and the zero-temperature bath-coupling spectrum 
\be
G_0(\omega) = \epsilon^2 \sum_k \eta^2_k(\hbar\omega_a)^2/\eta^2_{max} \delta(\omega-\omega_k).
\ee
\end{subequations}
The equilibrium value purity of the TLS is
\be
\label{eq::P_Eq}
P_{Eq} = \tanh(-\beta \hbar \omega_a/2),
\ee
with the ground and excited populations respectively given by 
\begin{align}
\rho_{ee}=&P_{Eq}^+=\left(1+P_{Eq}\right)/2,\\
\rho_{gg}=&P_{Eq}^-=\left(1-P_{Eq}\right)/2.
\end{align}
The frequency --- and temperature --- dependent coefficients in (\ref{hp}) are
\begin{multline}
\mathcal{K}^{\pm}(\omega) \
= \frac{1}{(1\mp\frac{\omega}{\omega_a})^2}\left[\left(\cosh \beta\hbar(\omega_a\mp\omega)-1\right)\right.\\
\left.\pm \left(\sinh \beta\hbar(\omega_a\mp\omega)-\beta\hbar(\omega_a\mp\omega)\right) \right].
\end{multline}

\begin{figure}[ht]
\includegraphics[width=1.\linewidth]{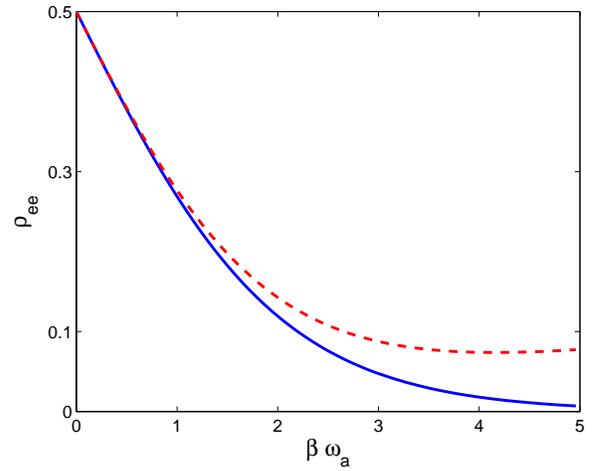}
\caption{\label{fig-purity} 
Excitation of the TLS at thermal equilibrium as a function of the
inverse temperature $\beta$ either with (dashed) and without (solid)
considering the effect of the system-bath interaction. Parameters:
memory time of the bath $t_c=2/\omega_a$, peak of the Lorentzian bath
spectrum $\omega_0=2\omega_a$, maximal coupling strength to the bath
$|\eta_{max}|^2=\omega_a/100$, where $\hbar\omega_a$ is the energy
separation of the TLS.}
\end{figure}

From Eq.~\eqref{hp} it can be seen that even at zero temperature
purity is incomplete, $P_{Eq}(\epsilon) < 1$, which is due to the
system-bath entanglement. The difference between $P_{Eq}$ in
(\ref{eq::P_Eq}) and $P_{Eq}(\epsilon)$ in (\ref{hp}) has a
non-monotonic dependence on $\beta$. This can be seen from
Fig.~\ref{fig-purity} where we have plotted the relative change of TLS
purity with inverse temperature. As the purity drop that we wish to
correct is non-monotonic with temperature, so will be the resultant
purification.

Using a similar analysis, the mean interaction energy to
$O(\epsilon^2)$, is given by
\begin{subequations}
\begin{multline}
\label{eq::H-SB-eps}
\langle H_{SB} (\epsilon) \rangle_{Eq}=-\hbar \omega_a \epsilon^2\\
\cdot\frac{\int^\infty_{-\infty}d\omega G_T(\omega)\left[P^+_{eq}\tilde{\mathcal{K}}^+(\omega) - P^-_{eq}\tilde{\mathcal{K}}^-(\omega) \right]} {1+\epsilon^2\int^\infty_{-\infty}d\omega G_T(\omega)\left[P^+_{eq}\mathcal{K}^+(\omega) + P^-_{eq}\mathcal{K}^-(\omega) \right]},
\end{multline}
where the quantity in brackets is dimensionless, and
\begin{multline}
\tilde{\mathcal{K}}^\pm = \frac{1}{1\mp\frac{\omega}{\omega_a}}\\
\cdot\big[ \cosh \beta \hbar(\omega_a\mp\omega)-1  \pm \sinh \beta\hbar (\omega_a\mp\omega)\big].
\end{multline}
\end{subequations}

For a Lorentzian coupling spectrum,
\begin{subequations}
\be
\eta_k=\eta_{max}\sqrt{\frac{\Gamma^2}{\Gamma^2+(\omega_0-\omega_k)^2}},
\ee
the mean interaction-energy at $T=0K$, is simply given by the
bath-induced lamb shift \cite{coh92}
\be
\langle H_{SB} \rangle_{Eq} \approx -\hbar \omega_a \int^\infty_0 d\omega\frac{\Gamma^2}{\Gamma^2 + (\omega_0 - \omega)^2} \frac{1}{(1+\omega/\omega_a)}.
\ee
\end{subequations}
This proves the {\em negativity} of the mean system-bath interaction
energy in equilibrium.

\section{Disturbance of equilibrium by impulsive QND measurement}
\label{sec-disturbance}
The Hamiltonian is intermittently perturbed by the coupling of the
system (qubit) to the detector (measuring apparatus), designed to
effect a QND impulsive measurement in the $\sigma_z$-basis. Such a
measurement projects the qubit onto the $\ket e$ or $\ket g$ energy
states. We stress that the measurement results are {\em unread}, i.e., the
qubit dynamics is changed by {\em non-selective measurements}.

\subsection{Dynamic description of the measurement}
\label{Sec::supp-A}
The time-dependent system-detector coupling (to the $k$th detector) has the form
\be \label{H-DS}
H_{SD}(t)=\frac{h(t)}{2} (1+\sigma_z)
\left(\ket{0}\bra{0} + \ket{1}\bra{1}-\ket{0}\bra{1}-\ket{1}\bra{0}\right).
\ee
where $(1+\sigma_z)=\ket{e}\bra{e}$ ensures QND measurement of the
qubit energy, and
\be
h(t)= \frac{\pi}{4\tau} \left({\rm
tanh}^2\left(\frac{t-t_0}{\tau}\right) - 1\right) \label{eq7}
\ee
is a smooth temporal profile of the system coupling to the detector qubits during 
the measurement that occurs at time $t_0$ and has a duration of $\tau$.

The detector (ancilla) qubits have energy-degenerate states $|0\rangle, |1\rangle$ so that
we may set the detector Hamiltonian to be zero 
\be 
H_D=0.
\ee

This form of the single-measurement Hamiltonian $H_{SD}$ was chosen so that the measurement interval is $[0,\tau]$:

\be
e^{-i\int_0^\tau dt H_{SD}(t)/\hbar} = U_C \label{eq8}.
\ee
where $U_C$ denotes to the CNOT operation.

In our model (Eqs. \eqref{H-I} - \eqref{eq8})
\begin{multline}
e^{-i\int_0^\tau dt H_{SD}(t)}\ket{0}_D = U_C|0\rangle_D\\
 = \ket{1}_D \ket{e}\bra{e} + \ket{0}_D \ket{g}\bra{g}.
\label{e37}
\end{multline}

The measurement consists in letting the TLS interact with the detector
(a degenerate TLS) via $H_{SD}$. The measurement outcomes are averaged
over (for non-selective measurements), by tracing out the detector
degree of freedom. The total effect on the system density-operator is:
\begin{multline}
\rho_S \mapsto Tr_D \left\{U_C\rho_S \otimes \ket{0}_{DD}\bra{0}\right\}\\
=\ket{e}\bra{e}\rho_S\ket{e}\bra{e}+\ket{g}\bra{g}\rho_S\ket{g}\bra{g} 
\end{multline}
i.e., the diagonal elements are unchanged, and the off-diagonals are
erased.  Since the TLS is entangled with the bath, the effect of the
measurement in Eq. \eqref{eq8} is:
\begin{align}
\rho_{tot}(0)=&\rho_{Eq}\rightarrow\rho_{tot}^M=Tr_D \left\{U_C\rho_{tot} \otimes \ket{0}_{DD}\bra{0} \right\}\nonumber\\
=&\ket{e}\bra{e}\rho_{tot}\ket{e}\bra{e}+
\ket{g}\bra{g}\rho_{tot}\ket{g}\bra{g}\nonumber\\ 
\equiv&\rho^B_{ee}\ket{e}\bra{e} + \rho^B_{gg}\ket{g}\bra{g}. \label{eq11}
\end{align}

Since $H_{SD}$ in Eq. \eqref{H-DS} commutes with $H_S$, we may
consider the measurement-induced evolution of $\mean{H_{SB}(\tau)}$,
rather than $\mean{H_{tot}(\tau)}$.  In the impulsive limit ($\tau
\rightarrow 0$), the measurement yields:
\begin{multline}
\mean{H_{SB}}_{EQ}\rightarrow\mean{H_{SB}(\tau)}^M\\
 =Tr \left\{ \rho_{tot}(0) \phantom{\rangle}_D\bra{0} U_{C}^\dagger H_{SB}(0) U_{C} \ket{0}_D  \right\}.
\end{multline}
Finally, using the RHS of \eqref{e37} and \eqref{H-I}, we get:
\begin{multline}
\phantom{\rangle}_D\bra{0} U_{C}^\dagger H_{SB}(0) U_{C} \ket{0}_D =0\\  \rightarrow  \mean{H_{SB}(\tau)}^M = 0.
\end{multline}

In fact, this result follows immediately from the nature of the
projective measurement:
\be
\begin{split}
\langle H_{SB} \rangle^M =& \frac{1}{2}\langle H_{SB} \rangle_{Eq} \\
&+\frac{1}{2}\sum_k \eta_k {\rm Tr}[(b_k+b^\dagger_k)   \sigma_z \sigma_x \sigma_z \rho_{Eq}] \\
=&\frac{1}{2}\langle H_{SB} \rangle_{Eq} - \frac{1}{2}\langle H_{SB} \rangle_{Eq} = 0,
\end{split}
\ee
where we have used the identity $\sigma_z \sigma_x \sigma_z = -\sigma_x$. 

This expresses the vanishing of
$Tr\left\{\rho_{tot}(\tau)H_{SB}\right\}^M$ due to the diagonality of
$\rho_{tot}^M(\tau)$ with respect to $S$. Since $H_D=0$, the detector
mean energy is not affected by the measurement.


\subsection{Post-measurement heating}


As shown in (\ref{Sec::supp-A}) above, a nearly-impulsive (projective)
quantum measurement ($\tau\rightarrow0$) of $S$, in the
$|g\rangle,~|e\rangle$ basis, using the energy supplied by
$H_{SD}(0<t<\tau)$ eliminates the mean system-bath interaction
energy. Now the pre-measurement equilibrium mean value,
$\mean{H_{SB}}_{Eq}$, is {\em negative}, as is shown above
(Eq.~(\ref{eq::H-SB-eps})) by second-order perturbation theory,
provided the temperature is positive, i.e., the $\ket{g}$ state is
populated more than the $\ket{e}$ state at thermal equilibrium. Hence
\begin{equation}
\begin{split}
&\mean{H_{SB}(0)}_{Eq}<0\mapsto\mean{H_{SB}(\tau)}^M=0,\\
&\mean{H_{SD}(t)}=-\mean{H_{SB}(t)}^M.
\end{split}
\label{eq:H-SB-H_SD}
\end{equation}

\begin{figure}[ht]
\includegraphics[width=1.\linewidth]{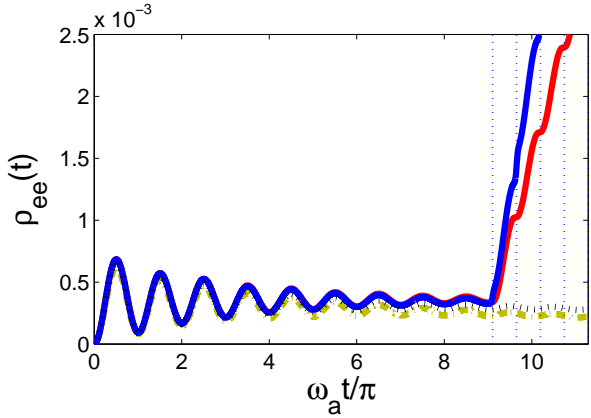}
\caption{\label{fig1} System evolution as a function of time. 
Excited-level population as a function of time for initially
zero-temperature product state, followed by relaxation to
quasi-equilibrium and then subjected to a series of measurements
(vertical dashed lines). Measurements of finite duration
($\tau_k=0.11/\omega_a$) (blue line) results in somewhat larger
heat-up than impulsive measurements (red line), but the dominant
effect is the same for both. Observe the agreement between $2^{nd}$
order master equation (green), two-quanta exchange with a discrete
bath, and exact numerical solution for a discrete bath of $40$ modes
(black dashed). Parameters: memory time of the bath $t_c=10/\omega_a$,
peak of the bath spectrum $\omega_0=\omega_a$, maximal coupling
strength to the bath $|\eta_{max}|^2=0.07\omega_a$, where
$\hbar\omega_a$ is the energy separation of the TLS.}
\end{figure}

\begin{figure}[ht]
\includegraphics[width=1.\linewidth]{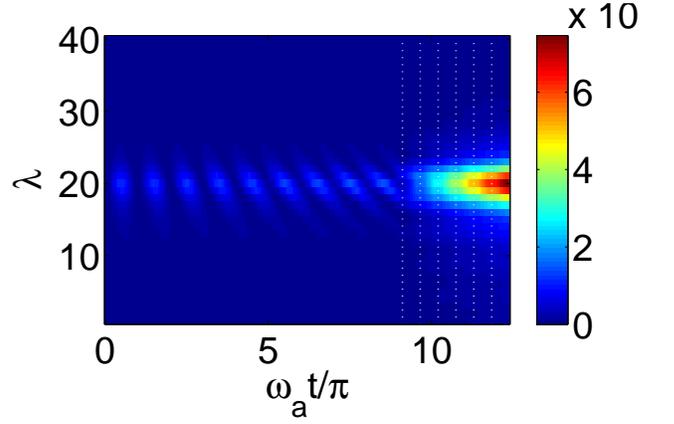}
\caption{\label{fig1c}
Excitations as a function of $t$ of the $40$ modes in the two-quanta
model. $\lambda$ is the mode number. Parameters as in Fig.~\ref{fig1}
}
\end{figure}

After the measurement (as $H_{SD}(t\ge\tau)=0$), time-energy
uncertainty at $\Delta t \lesssim 1/\omega_a$ results in the breakdown
of the RWA, i.e., $\langle H_S + H_B \rangle$ is not conserved as
$\Delta t$ grows. The resulting $\mean{H_S}+\mean{H_B}$ changes stem
from the non-commutativity of $H_{SB}$ and $H_{SD}$.  Only $\langle
H_{tot} \rangle$ is conserved, by unitarity, until the next
measurement. Hence, the {\em post-measurement decrease} of $\langle
H_{SB} \rangle$ with $\Delta t$, signifying the restoration of
equilibrium:
\be
\langle H_{SB}(\tau)\rangle^M =0
\rightarrow \langle H_{SB}(\tau+\Delta t)\rangle < 0,
\ee
is at the expense of the \emph{increase} 
\be
\langle H_S + H_B\rangle =
\langle H_{tot}\rangle - \langle H_{SB}\rangle>0,
\ee
i.e., \emph{heating} of the system and the bath (Fig.~\ref{fig1},
~\ref{fig1c}), combined.

\subsection{Short-time post-measurement qubit evolution}
Let us denote the even part of the bath state by $|B^{{\rm
even}}\rangle|e\rangle$ and that of the odd part as $|B^{{\rm
odd}}\rangle$, then:

\begin{equation}
|B^{{\rm
even}}_{\bm{n},g}(t)\rangle\otimes|g\rangle+|B^{{\rm
odd}}_{\bm{n},g}(t)\rangle\otimes|e\rangle\equiv |\Psi_{\bm{n},g}
(t)\rangle. \label{eq28}
\end{equation}
Here $B^{{\rm even}}$ (respectively, $B^{{\rm odd}}$) is a combination
of bath $\hat{N}$-eigenstates with eigenvalues differing from
$\hat{N}$ by even (respectively, odd) numbers.

The post-measurement evolution of the system alone, described by
$\rho_S=Tr_B\rho_{tot}$, is not at all obvious. Its Taylor expansion holds at short
evolution times, $\Delta t \ll 1/\omega_a$,
\be
\label{rho-s-taylor}
\rho_S(\tau+\Delta t)\simeq \rho_S(\tau)+\Delta
t\dot\rho_S(\tau)+\frac{\Delta t^2}{2}\ddot{\rho}_S(\tau)+\ldots
\ee
The $0$th order term is {\em unchanged} by the measurement,
$\rho_S(\tau)=\rho_S(t\le0)$. 

Due to the post-measurement vanishing of the off-diagonal elements of $\rho_{tot}$ (Eq. (\ref{eq11}),
for $\rho_{tot}(t) =  |\Psi_{\bm{n},g} (t)\rangle  \langle \Psi_{\bm{n},g} (t)|$ (Eq. \eqref{eq28}), we have 
\begin{equation}
\begin{split}
\left(\rho_S\right)_{eg}(t)=&\bra{e}\rho_S(t)\ket{g}=Tr_{B}\langle e | \rho_{tot}(t)| g\rangle\\
 =& \langle B^{{\rm
even}}_{\bm{n},e}(t	)|B^{{\rm  odd}}_{\bm{n},e}(t)\rangle=0 \label{e24}
\end{split}
\end{equation}
Hence, $\rho_S$ is diagonal at any time $t$. 

Its derivative immediately after the measurement,
$\dot{\rho_S}(\tau)$, has the form:
\begin{align}
\dot{\rho_{S}}(\tau)=&
-ie^{-i\omega_{a}\tau}\ket{e}\bra{g}
Tr_{B} \left\{\mathcal{B}\left(\rho_{gg}^{B}-\rho_{ee}^{B}\right)\right\}
+ H.C.\nonumber\\
=&0.
\label{eq35}
\end{align}

The same argument goes through upon permuting $e \leftrightarrow g$ everywhere for $\rho_{tot} =  |\Psi_{\bm{n},e} (t)\rangle  \langle \Psi_{\bm{n},e} (t)|$.

Hence, the first derivative {\em vanishes} at $t=\tau (\Delta t=0)$
due to the definite parity of the bath density-operator correlated to
$\ket{g}$ or $\ket{e}$. This post-measurement vanishing of the first
derivative, $\dot{\rho}_S(\tau)=0$, is the condition for the quantum
Zeno effect (QZE)\cite{mis77,kof00,kof04,facchi2001po}. The time
evolution of $\rho_S$ is then governed by its second time derivative
$\ddot{\rho}_{S}(\tau)$.

For the {\em factorisable} thermal state, 
\be
\rho_{tot}=Z^{-1}e^{-\beta H_0}=Z_B^{-1}e^{-\beta H_B} Z_S^{-1}e^{-\beta H_S}, 
\ee
we have:
\begin{equation}
\label{eq32}
\begin{split}
\rho^B_{ee} \equiv& \langle e | \rho_{tot}| e\rangle = \langle e
|Z_S^{-1}e^{-\beta H_S} | e \rangle Z_B^{-1}e^{-\beta H_B}\\
=&\left(\rho_S\right)_{ee} \rho_B ~({\rm and~} e \leftrightarrow g ).
\end{split}
\end{equation}

For this $\rho_{tot}$, the second derivative of $\rho_S$ immediately after the measurement is (cf. Eq. \eqref{eq11})
\be
\ddot{\rho}_{S}(\tau)=2\sigma_z Tr_{B}\left\{
\mathcal{B}^2(\rho_{gg}^{B}-\rho_{ee}^{B})\right\}.
\label{eq14}
\ee
The scalar factor is positive:
\begin{multline}
Tr_{B}\left\{\hat{\mathcal{B}}^2\left(\rho^B_{gg}-\rho^B_{ee}\right)
\right\}\\
 =Tr_{B} \left\{\hat{\mathcal{B}}^2 \rho_B
\right\}\left( \left(\rho_S\right)_{gg}-\left(\rho_S\right)_{ee}\right)
> 0, 
\label{eq47}
\end{multline} 
where we have used $Tr_B\{\rho^B_{gg(ee)}\}=\left(\rho_S\right)_{gg(ee)}$ which follows from the definition (Eq.\eqref{eq11}):
$\rho^B_{ee(gg)} = \bra{e(g)}\rho_{tot}\ket{e(g)}$.
The first factor in \eqref{eq47} is positive by virtue of the positivity of the operator $\hat{\mathcal{B}}^2$
($\hat{\mathcal{B}}$ being Hermitian), and the second is
positive iff there is no population inversion for the TLS. 

Hence, the second derivative in \eqref{rho-s-taylor} is
\emph{positive} shortly after the measurement, if there is no initial
population inversion of the system, i.e., for non-negative
temperature.

\subsection{Post-measurement state}
The combined (system- and bath-) equilibrium state satisfies: 
\be
\rho_{tot}^M = Z^{-1} e^{-\beta H_{tot}} = \rho_{tot}=Z^{-1}e^{-\beta \left( H_0 + O(H_{SB}^2) \right)}.
\ee 
Thus, for sufficiently weak coupling, Eq. \eqref{eq32} dominates.

How is this reconciled with the non-unitary nature of the projection,
whereby the mixedness of the total state must increase? Indeed,
\be
{\rm Tr}[(\rho^M)^2] = \frac{1}{2}{\rm Tr}[(\rho_{Eq})^2]+\frac{1}{2}{\rm Tr}[(\sigma_z\rho_{Eq}\sigma_z\rho_{Eq})] 
\ee
Yet, in the weak-coupling limit, the increase in mixedness due to measurement is  $\cong O(\epsilon^4)$ and hence can be neglected.

\section{Post-measurement free evolution of the qubit}
\label{sec-main-text}
The evolution of $\rho_S$ at longer times (in the regime of weak
system-bath coupling) may be approximately described (as verified by
our {\em exact} numerical simulations\cite{nest2003dqd}) by the
second-order non-Markovian master equation (ME)\cite{breuer2002toq}
(Fig.~\ref{fig1}). Higher-order corrections to the ME
will be discussed elsewhere. The 2nd order ME for $\rho_S$, on account
of its diagonality, can be cast into the following population rate
equations\cite{kof04}, dropping the subscript $S$ in what follows and
setting the measurement time to be $t=0$:
\begin{align}
\dot\rho_{ee}(t) =& -\dot\rho_{gg}(t) =
R_g(t)\rho_{gg}-R_e(t)\rho_{ee},\\
\label{R-def}
R_{e(g)}(t) =&  2\pi t\int_{-\infty}^\infty d\omega
G_T(\omega)\sinc\left[(\omega\mp\omega_a)t\right].
\end{align}
Here $\sinc(x)=\frac{\sin(x)}{x}$. We shall assume that $G_0(\omega)$,
the zero-temperature coupling spectrum, has peak coupling strength at
$\omega_0$ and spectral width $\sim 1/t_c$.

\begin{figure}[ht]
\includegraphics[width=1.\linewidth]{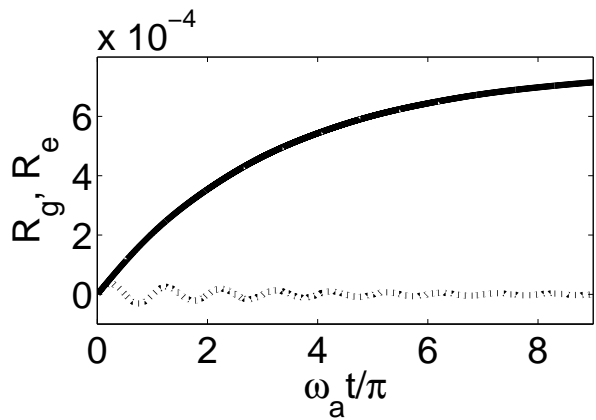}
\caption{\label{fig1a}
Relaxation rates, $R_g$ (dashed) and $R_e$ (solid) as a function of time.
Parameters as in Fig.~\ref{fig1}.}
\end{figure}

 \begin{figure}[ht]
\includegraphics[width=1.\linewidth]{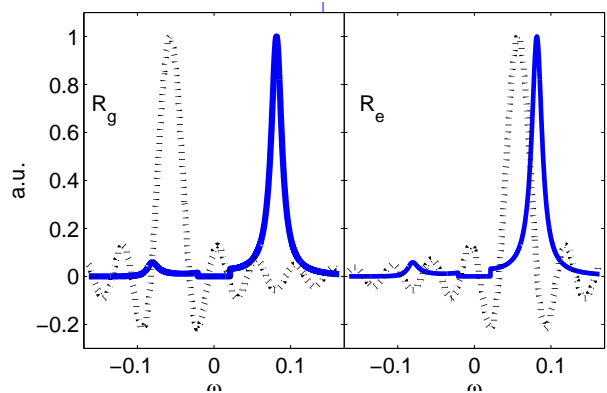}
\caption{\label{fig2a} $R_g(t)$ and $R_e(t)$ (Eq.~\eqref{R-def}) depicted as spectral overlaps of coupling spectrum (blue solid) and $\sinc((\omega\pm\omega_a)t)$ (black dashed). }
\end{figure}

The entire dynamics is determined by $R_{e(g)}(t)$ (Figs.~\ref{fig1a},~\ref{fig2a}), the relaxation
rates of the excited (ground) states: 

(i) At short times
$t\ll  1/\omega_a \ll t_c$ the $\sinc$ function in (\ref{R-def}) is
much broader than $G_T$. The relaxation rates $R_e$ and $R_g$ are
then equal at any temperature, indicating the complete breakdown of the RWA
discussed above: $|g\rangle \rightarrow |e\rangle$ and $|e\rangle
\rightarrow |g\rangle$ transitions do not require quantum
absorption or emission by the bath, respectively. The rates
$R_{e(g)}$ then become {\em linear} in time, manifesting the
QZE\cite{kof00,facchi2001po,kof04}:
\begin{align}
\label{R-short}
&R_{e(g)}(t\ll t_c)\approx 2 \dot R_0 t,\\
\label{Rdot-def}
&\dot{R}_0\equiv\int_{-\infty}^\infty d\omega G_T(\omega)  =
\mean{\mathcal{B}^2}.
\end{align}
This short-time regime entails the {\em universal Zeno heating rate}:
\be
\frac{d}{dt}\left(\rho_{ee}-\rho_{gg}\right) \approx 4 \dot{R}_0 t(\rho_{gg}-\rho_{ee}).
\ee

(ii) At intermediate non-Markovian times, $t \sim 1/\omega_a$,
when the $\sinc$ function and $G_T$ in \r{R-def} have comparable
widths, the relaxation rates $R_{e(g)}(t)$ exhibit several unusual
phenomena that stem from time-energy uncertainty. The change in the overlap of the $\sinc$ and $G_T$ functions with time results
in damped aperiodic oscillations of $R_e(t)$ and $R_g(t)$, near
the frequencies $\omega_0-\omega_a$ and $\omega_0+\omega_a$,
respectively. This oscillatory time dependence that conforms
neither to QZE nor to the converse AZE of relaxation speedup\cite{lan83,kof00,facchi2001po}, will henceforth be dubbed the {\em
oscillatory Zeno effect} (OZE). Due to the negativity of the
$\sinc$ function between its consecutive maxima, we can have a {\em negative relaxation rate}, 
which is completely forbidden by the RWA. Since $\sinc\left[(\omega+\omega_a)t
\right]$ is much further shifted from the peak of $G_T(\omega)$ than
$\sinc\left[(\omega -\omega_a)t \right]$, $R_g(t)$ is more likely
to be negative than $R_e(t)$ (Figs.~\ref{fig1a},~\ref{fig2a}). Hence,
$\rho_{gg}(t)$ may grow at the expense of $\rho_{ee}(t)$ more than allowed by the thermal-equilibrium detailed balance. This may cause {\em transient cooling}, as detailed below.

(iii) At long times
$t\gg t_c$, the relaxation rates attain their Golden-Rule (Markov)
values\cite{kof04}
\be
\label{R-long}
R_{e(g)}(t\gg t_c) \simeq 2\pi G_T(\pm\omega_a).
\ee
The populations then approach those of an equilibrium Gibbs state whose
temperature is equal to that of the thermal bath (Fig.~\ref{fig1}).

\begin{figure}[ht]
\includegraphics[width=1.\linewidth]{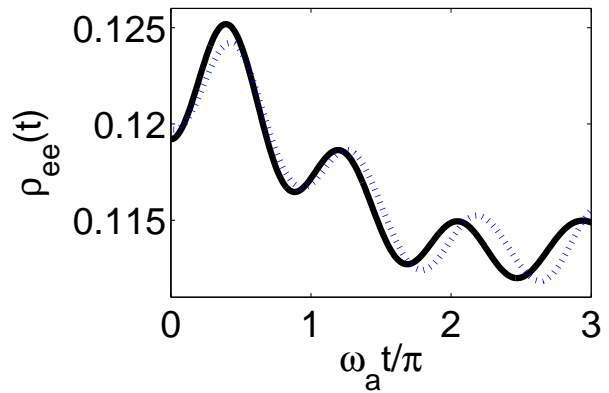}
\caption{\label{fig2b} Example of a system experiencing first Zeno
 heating, then \emph{oscillatory}-Zeno cooling, obtained from the
second-order master equation (black-solid) and from the exact numerical solution
for a discrete bath of 40 modes (blue dashed). }
\end{figure}

\begin{figure}[ht]
\includegraphics[width=1.\linewidth]{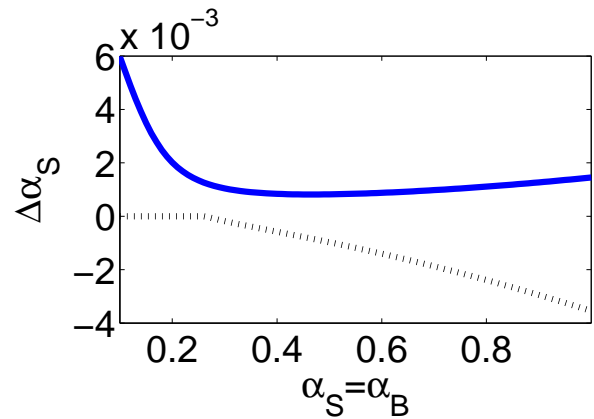}
\caption{\label{fig2c} Maximal Zeno heating (blue solid)
 and subsequent maximal cooling (black dashed) as a function of common
 initial temperature of system and bath
 $\alpha_S=\alpha_B=\hbar\omega_a\beta$. Note the {\em critical
 temperature} for oscillatory-Zeno cooling. Parameters: memory time of
 the bath $t_c=10/\omega_a$, peak of the bath spectrum
 $\omega_0=\omega_a/0.7$, maximal coupling strength to the bath
 $|\eta_{max}|^2=4.36\omega_a$. These effects can be {\em strongly
 magnified} by choosing other suitable parameters. }
\end{figure}

If we repeat this procedure often enough, the TLS will either
increasingly heat up or cool down, upon choosing the time
intervals $\Delta t_k$ to coincide with either peaks or troughs of the
$\rho_{ee}$ oscillations, respectively. Since consecutive measurements affect the
bath and the system differently, they may acquire different
temperatures, which then become the initial conditions for
subsequent QZE heating or OZE cooling, Fig.~\ref{fig2b}. The results are shown 
for both different and common (Fig.~\ref{fig2c})
temperatures of the system and the bath. Remarkably, the system may heat up solely
due to the QZE, although the {\em bath is colder}, or cool down solely
due to the OZE or AZE, although the {\em bath is hotter}. The
bath may undergo changes in temperature and entropy too (Fig.~\ref{fig1c}).

\section{Derivation of cooling conditions}
\label{sec-cooling-cond}
By integrating Eq.~\eqref{R-def} over time to acquire $J_{g(e)}(t)$, one arrives at the following result:
\begin{multline}
J_g(t)=\int_0^\infty d\omega G_0(\omega)n_T(\omega)\sinc^2((\omega-\omega_a)t)\\
+\int_0^\infty d\omega G_0(\omega)(n_T(\omega)+1)\sinc^2((\omega+\omega_a)t)
\end{multline}
\begin{multline}
J_e(t)=\int_0^\infty d\omega G_0(\omega)n_T(\omega)\sinc^2((\omega+\omega_a)t)\\
+\int_0^\infty d\omega G_0(\omega)(n_T(\omega)+1)\sinc^2((\omega-\omega_a)t)
\end{multline}
To obtain cooling below the equilibrium temperature, one requires that:
\be
\frac{J_e(t)}{J_g(t)}>\frac{J_e^{Eq}}{J_g^{Eq}} = \frac{R_e^{Eq}}{R_g^{Eq}} = \frac{n(\omega_a)}{n(\omega_a)+1}
\ee
Rearranging the terms in the above equation, gives the cooling condition,
\begin{equation}
\begin{split}
\label{eq:coolcond}
&\int_{0}^{\infty}G_{0}(\omega)\frac{\sin^{2}\left[\frac{t}{2}(\omega-\omega_{a})\right]}{(\omega-\omega_{a})^{2}}(n_T(\omega_{a})-n_T(\omega))
\\
>&\int_{0}^{\infty}G_{0}(\omega)\frac{\sin^{2}\left[\frac{t}{2}(\omega+\omega_{a})\right]}{(\omega+\omega_{a})^{2}}(n_T(\omega_{a})+n_T(\omega)+1)
\end{split}
\end{equation}

A general quest for finding the spectral density function
$G_0(\omega)$,which satisfies the above condition in some time
interval, at any given temperature $T$, is quite difficult. In the
high temperature limit i.e., $n_T(\omega) >> 1$ one can find a
necessary condition on the peak position of $G_0(\omega)$, which can
satisfy the above inequality.  Substituting the high-temperature limit
for $n_T(\omega)\simeq 1/\beta\omega$, one can show that in order to
allow cooling $G_0(\omega)$ needs to be concentrated in the frequency
interval defined by
\be
\omega_a < \omega < \Omega,
\ee
where
\be
\beta\Omega = 1 + \frac{\beta\omega_a + \sqrt{4+12\beta\omega_a+\beta^2\omega_a^2}}{2}.
\ee
Though $\beta\Omega >1$, it is only the maximum possible bound on the
detuning of the bath spectrum for the qubit frequency, indicating that
one should not detune the bath spectrum too far from $\omega_a$ to see
the cooling effect. We have numerically verified these conditions for
various bath coupling spectrums. In the same spirit one can find
regions in frequency space, where for specific times there will be no
cooling, independent of the shape of $G_{0}(\omega)$.

\section{Entropy dynamics}
\label{sec-supp-D}

One may always \emph{define} the entropy of $\rho_S$ \emph{relative} to its
equilibrium state $\rho_0$ (``entropy distance'') and the negative
of its rate of change, as\cite{ali79,lin74}:

\begin{multline}
\bm{S}(\rho_S(t)||\rho_0)\\
\equiv\mathrm{Tr}\{\rho_S(t) \ln \rho_S(t)\}-\mathrm{Tr}\{\rho_S(t) \ln \rho_0\}
\end{multline}
\begin{equation}
\bm{\sigma}(t) \equiv-\frac{d}{dt} \bm{S}(\rho_S(t)||\rho_0).
\label{sigma-def}
\end{equation}

\begin{figure}[ht]
\includegraphics[width=1.\linewidth]{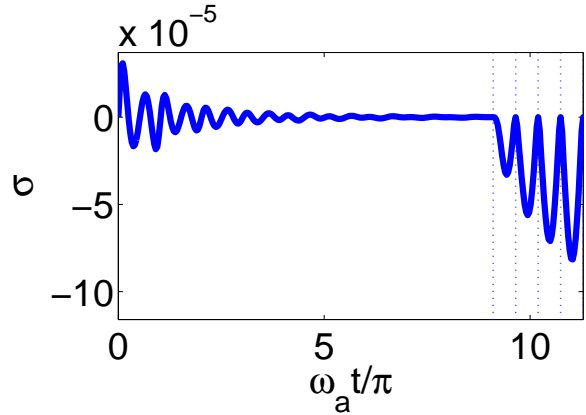}
\caption{\label{fig1b}
${\bm\sigma}(t)$ (negative of relative entropy rate of change). 
Parameters as in Fig.~\ref{fig1}.}
\end{figure}

\emph{In the Markovian realm} $\bm{\sigma}(t) \geq 0$\cite{spo78,ali79,lin74} is a
statement of the second law of thermodynamics. Since $\rho_S$ is
diagonal, it follows that $\bm{\sigma(t)}$ is
positive iff
$\frac{d}{dt}\left|\rho_{ee}(t)-(\rho_0)_{ee}\right|\leq 0$,
consistently with the interpretation of the relative entropy
$\bm{S}(\rho_S||\rho_0)$ in (\ref{sigma-def}) as the entropic
``distance'' from equilibrium. Conversely, whenever the
oscillatory $\rho_{ee}(t)$ drifts away from its initial or final
equilibria, $\bm{\sigma}$ takes negative values (Fig.~\ref{fig1b}).

\section{Discussion: realization and practical consequences}
\label{sec-conc}
\begin{figure}[ht]
  \centering
\mbox{
\subfigure[Atoms in microwave cavity]
{\includegraphics[width=0.5\linewidth]{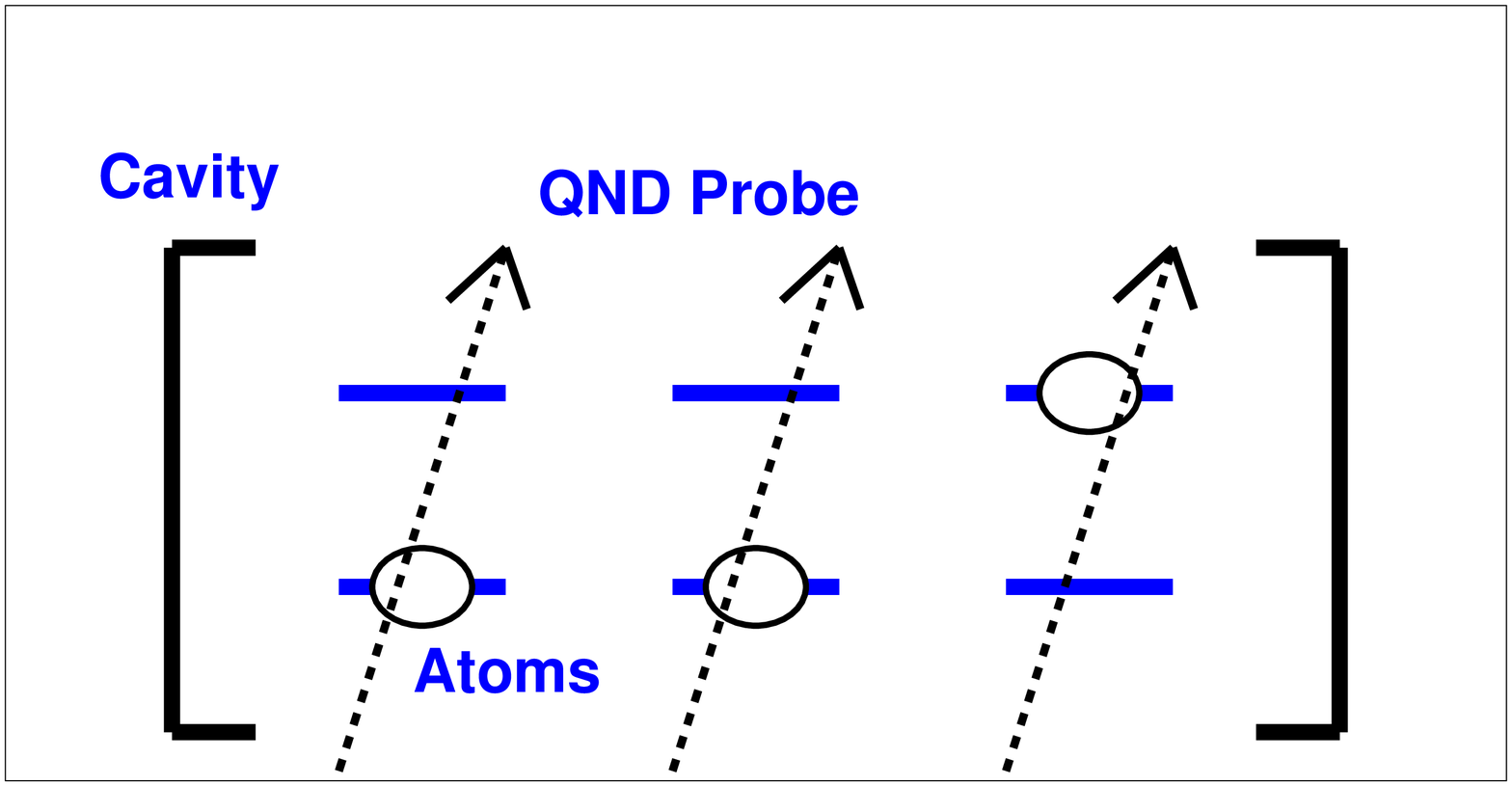}\label{fig2d}}
\subfigure[Energy level description]
{\includegraphics[width=0.4\linewidth]{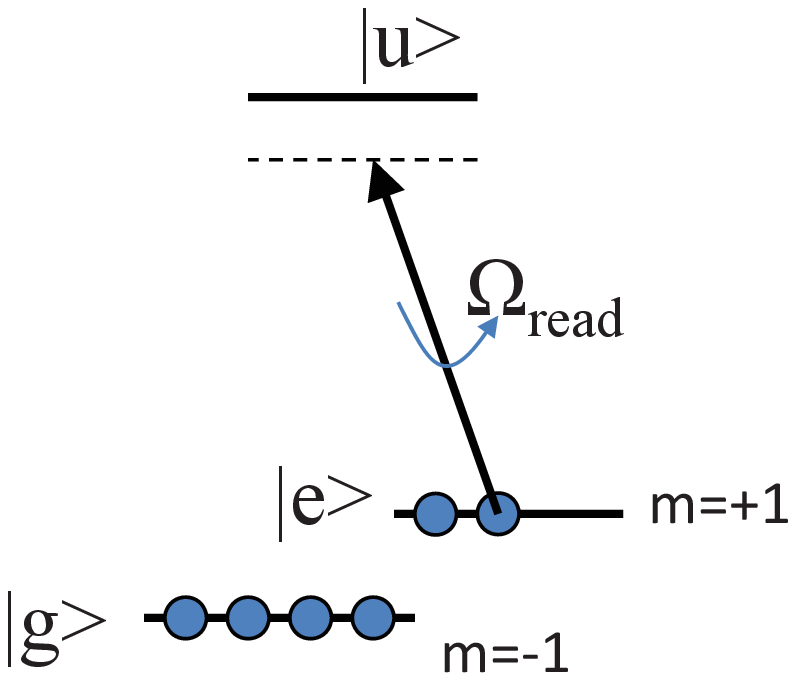}\label{fig10}}
}
\caption{Possible experimental setup}
  \label{fig::exper_setup}
\end{figure}

Consider atoms or molecules in a microwave cavity (Fig.~\ref{fig2d}) with controllable
finite-temperature coupling spectrum $G_T(\omega)$ centered at
$\omega_0$. Measurements can be effected on such a TLS
ensemble with resonance frequency $\omega_a$ in the microwave
domain, at time intervals $\Delta t_k \sim 1/(\omega_0 \pm
\omega_a)$, by an optical QND probe\cite{braginsky1995qm} at
frequency $\omega_p \gg\omega_a, \omega_0$. The probe pulses
undergo different Kerr-nonlinear phase shifts $\Delta \phi_{e}$ or
$\Delta \phi_{g}$ depending on the different symmetries (e.g.,
angular momenta) of $|e\rangle$ and $|g\rangle$. The relative
abundance of $\Delta \phi_e$ and $\Delta \phi_g$ would then
reflect the ratio $\rho_{ee}(t_k)/\rho_{gg}(t_k)$. Such QND
probing may be performed with time-duration {\em much shorter than}
$\omega_a^{-1}$, i.e. $\omega_a \tau_k\ll 1$, {\em without resolving
the energies} of $|e\rangle$ and  $|g\rangle$.

An experimental scenario involves collective N-atom coupling to near-resonant RF resonator (Fig.~\ref{fig2d}).
Let us choose ground sublevels $\ket{m=-1}\equiv\ket{g}, \ket{m=+1}\equiv\ket{e}$, with Zeeman splitting $\omega_{eg}\cong MHz$.
The collective Rabi frequency of $N\ge 10^8$ atoms at a cavity antinode is $N^{1/2}\Omega_{eg}/2\pi\cong 100KHz$.
An optical beam will rotate in polarization (Fig.~\ref{fig10}), thus performing QND measurement (readout) that resolves $\ket{g}$ and $\ket{e}$ (by their symmetry, not by energy) if its Rabi frequency:
\be
\Omega_{\rm read}\cong N^{1/2}(\Omega_0)_{read}^2/\Delta\ge\omega_{eg}.
\ee
Such a Rabi frequency corresponds to RWA violation, as discussed in the text above.


Non-selective measurements increase the Von-Neumann entropy
of the detector ancillae. Since our ancillae are laser pulses, they are only used once and we
may progressively change the TLS ensemble thermodynamics by consecutive
pulses, disregarding their entropic or energetic price. 

\section{Conclusions}
To conclude, we have shown that frequent QND measurements may induce
either anomalous heating or anomalous cooling of TLS coupled to baths
on non-Markovian time scales. These findings defy the standard notions
of quantum thermodynamics regarding system equilibration in the
presence of a thermal bath.

The practical advantage of the predicted anomalies is the
possibility of {\em very rapid control} of cooling and entropy, which
may be attained after several measurements at $t\geq\omega_a^{-1}$ and
is only limited by the measurement rate. By contrast, conventional
cooling requires much longer times, $t\gg t_c$, to reach thermal
equilibrium.

\section*{Acknowledgments}
We acknowledge the support of ISF, GIF and EC.


\end{document}